\documentclass[10pt,aps,prl,twocolumn,preprintnumbers,amsmath,amssymb,floatfix,showpacs,citeautoscript]{revtex4-1}
\usepackage[latin1]{inputenc}
\usepackage[T1]{fontenc}
\usepackage[english]{babel}
\usepackage[pdftex]{graphicx}
\usepackage{amsmath}
\usepackage{times}
\usepackage[usenames,dvipsnames,svgnames,table]{xcolor}
\usepackage[colorlinks,plainpages=false,linkcolor=blue,urlcolor=blue,citecolor=blue,pdfpagemode=UseNone]{hyperref}

\renewcommand{\AA}{\text{\r{A}}}

\newcommand\Vek[1]{\vec{#1}}

\newcommand\VO{V$_\text{O}$ }
\newcommand\VOs{V$_\text{O}$}
\newcommand\muB{\mu_{\text{B}}}

\newcommand\lc[1]{\lowercase{#1}}

\begin{document}

\title
{
\boldmath
Oxygen vacancy formation and electronic reconstruction in strained LaNiO$_3$ \\ and LaNiO$_3$/LaAlO$_3$ superlattices
}

\author{Benjamin Geisler}
\email{benjamin.geisler@uni-due.de}
\affiliation{Department of Physics and Center for Nanointegration (CENIDE), Universit\"at Duisburg-Essen, Lotharstr.~1, 47057 Duisburg, Germany}
\author{Simon Follmann}
\affiliation{Department of Physics and Center for Nanointegration (CENIDE), Universit\"at Duisburg-Essen, Lotharstr.~1, 47057 Duisburg, Germany}
\author{Rossitza Pentcheva}
\email{rossitza.pentcheva@uni-due.de}
\affiliation{Department of Physics and Center for Nanointegration (CENIDE), Universit\"at Duisburg-Essen, Lotharstr.~1, 47057 Duisburg, Germany}

\date{\today}

\begin{abstract}
By using density functional theory calculations including a Coulomb repulsion term,
we explore the formation of oxygen vacancies and their impact on the electronic and magnetic structure
in strained bulk LaNiO$_3$ and (LaNiO$_3$)$_1$/(LaAlO$_3$)$_1$(001) superlattices.
For bulk LaNiO$_3$, we find that epitaxial strain induces a substantial anisotropy
in the oxygen vacancy formation energy.
In particular, tensile strain promotes the selective reduction of apical oxygen,
which may explain why the recently observed superconductivity of infinite-layer nickelates is limited to strained films.
For (LaNiO$_3$)$_1$/(LaAlO$_3$)$_1$(001) superlattices,
the simulations reveal that the NiO$_2$ layer is most prone to vacancy formation,
whereas the AlO$_2$ layer exhibits generally the highest formation energies.
The reduction is consistently endothermic, 
and a largely repulsive vacancy-vacancy interaction is identified
as a function of the vacancy concentration. 
The released electrons are accommodated exclusively in the NiO$_2$ layer,
reducing the vacancy formation energy in the AlO$_2$ layer by $\sim\!70\%$ with respect to bulk LaAlO$_3$.
By varying the vacancy concentration from $0\%$ to $8.3\%$
in the NiO$_2$ layer at tensile strain,
we observe an unexpected transition from a localized site-disproportionated ($0.5\%$)
to a delocalized ($2.1\%$) charge accommodation,
a re-entrant site disproportionation leading to
a metal-to-insulator transition despite a half-filled majority-spin Ni~$e_g$ manifold ($4.2\%$), 
and finally a magnetic phase transition ($8.3\%$).
While a band gap of up to $0.5$~eV opens at $4.2\%$ for compressive strain,
it is smaller for tensile strain or the system is metallic,
which is in sharp contrast to the defect-free superlattice.
The strong interplay of electronic reconstructions and structural modifications
induced by oxygen vacancies in this system % correlated quantum matter
highlights the key role of an explicit supercell treatment beyond rigid-bands methods,
and exemplifies the complex response to defects in artificial transition metal oxides.
\end{abstract}

\maketitle

\section{Introduction}

\vspace*{-1em}

The recent observation of superconductivity in infinite-layer NdNiO$_2$~\cite{Li-Supercond-Inf-NNO-STO:19, Li-Supercond-Dome-Inf-NNO-STO:20}, PrNiO$_2$~\cite{Osada-PrNiO2-SC:20}, and LaNiO$_2$~\cite{Osada-LaNiO2-SC:21} films on SrTiO$_3$(001) (STO)
has re-initiated considerable interest in rare-earth nickelates~\cite{Nomura-Inf-NNO:19, JiangZhong-InfNickelates:19, Sakakibara:19, JiangBerciuSawatzky:19, Botana-Inf-Nickelates:19, Lechermann-Inf:20, NNO-SC-Thomale:20, NNO-SelfDopingDesign-d9-Arita:20, Kitatani-AritaZhongHeld:20, GeislerPentcheva-InfNNO:20, Choi-Lee-Pickett-4fNNO:20, Si-Zhonh-Held:InfNNO-Hydrogen:20, WangKangMiaoKotliar:20, Gu-NNO2:20, GeislerPentcheva-NNOCCOSTO:21, Ortiz-NNO:21, Lu-MagExNdNiO2:21, GoodgeGeisler-NNO-IF:22}.
However, the physical mechanism behind this observation remained elusive so far. % is so far not fully understood.
Specifically, it is unclear why superconductivity is absent in bulk samples~\cite{Li-NoSCinBulkDopedNNO:19, Wang-NoSCinBulkDopedNNO:20}.
The role of the polar interface to the STO(001) substrate and a thereby induced electronic reconstruction
towards a cuprate-like Fermi surface
has been pointed out recently~\cite{GeislerPentcheva-InfNNO:20, GeislerPentcheva-NNOCCOSTO:21, GoodgeGeisler-NNO-IF:22}.
A central step % during experimental sample fabrication 
in the reduction from the perovskite $AB$O$_3$ to the infinite-layer $AB$O$_2$ phase
is the targeted deintercalation of apical oxygen ions.
However, in rhombohedral bulk LaNiO$_3$ (LNO), all oxygen sites are symmetry-equivalent.
This is also valid in good approximation for the orthorhombic compounds PrNiO$_3$ and NdNiO$_3$.
A site-selective versus the expected statistical reduction may explain why superconductivity arises exclusively in film geometry.

The aspect of anisotropy brings another system back into focus:
(LaNiO$_3$)$_1$/(LaAlO$_3$)$_1$(001) (LNO/LAO) superlattices (SLs),
where the formation of a cuprate-like Fermi surface was proposed for a single layer of the correlated metal LNO 
confined in a SL with the band insulator LAO~\cite{ChaloupkaKhaliullin:08, Hansmann:09, Han-ChemCtrl:10}.
The single-band Fermi surface would arise as a result of a high  orbital polarization of the formally singly occupied Ni~$e_g$  states, which could not be confirmed in later studies~\cite{ABR:11, Boris:11, WuBenckiser:13}.
Instead, a site and bond disproportionation expressed in distinct magnetic moments and Ni-O bond lengths at adjacent Ni sites was found for (LNO)$_1$/(LAO)$_1$(001)~\cite{ABR:11, Freeland:11}. % under tensile strain % arises irresp of strain!
This bears similarities to bulk rare-earth nickelates, where
% a Ni$^{3+}$ $\to$ Ni$^{3+\delta}$ $+$ Ni$^{3-\delta}$
site disproportionation has been considered as the origin of the metal-to-insulator transition (MIT)~\cite{Mazin:07, ParkMillisMarianetti:12, JohnstonCD:14, Varignon:17}.
Related effects emerge in ultrathin nickelate films and heterostructures~\cite{RENickelateReview:16, WrobelGeisler:18}
as well as in cobaltates~\cite{Kunes-LCO-Disprop-Model:11, GeislerPentcheva-LCO:20}. % bulk and strained films
Moreover, LNO/LAO heterostructures~\cite{Benckiser:11, LiuChakhalian:11, Puggioni:12, Frano:13, LuBenckiser:16, LNOLAO-4-4-ParkMillisMarianetti:16}
show considerable potential for thermoelectric applications~\cite{Viewpoint:19, TE-Oxides-Review-Geisler:21},
either by exploiting the emergence of a $\sim 0.3$~eV band gap in (LNO)$_1$/(LAO)$_1$(001) SLs strained to the lattice constant of STO~\cite{GeislerPentcheva-LNOLAO:18}
or by strain-tuning of orbital-selective transport resonances in (LNO)$_3$/(LAO)$_1$(001) SLs~\cite{GeislerPentcheva-LNOLAO-Resonances:19}.
These aspects raise a question about the robustness of the electronic phase with respect to oxygen vacancies.

Here we explore the formation of oxygen vacancies and their impact on the electronic and magnetic structure
in strained bulk LNO and (LNO)$_1$/(LAO)$_1$(001) SLs
by performing density functional theory calculations including a Coulomb repulsion term.
The vacancies are treated explicitly via large supercells, which proves to be essential.
First, we show for the bulk LNO reference system
that epitaxial strain induces a substantial anisotropy in the oxygen vacancy formation energy.
Tensile strain as exerted by STO(001) enhances the formation of apical vacancies by $25$~meV per vacancy,
which facilitates a targeted reduction to the infinite-layer phase
and offers a possible explanation for the exclusive observation of superconductivity in strained infinite-layer nickelate films.
We find that the electrons released by the vacancies are predominantly accommodated locally despite the overall metallic phase,
irrespective of strain,
resulting in the collapse of the adjacent Ni magnetic moments. % at variance with Hund's rule.
Moreover, we identify bulk LNO at $4.2\%$ vacancy concentration to be at the verge of a vacancy-induced MIT.

Subsequently, we investigate the formation of oxygen vacancies in the (LNO)$_1$/(LAO)$_1$(001) system, % explicitly anisotropic 
particularity the layer and strain dependence of the reduction energies,
and elucidate different accommodation mechanisms of the released electrons.
Specifically, we probe the impact of oxygen vacancies on the site-disproportionated quantum phase,
systematically varying their concentration between $0\%$ and $8.3\%$.
We find that oxygen vacancies form preferably in the NiO$_2$ layers,
with a minimum formation energy of $2.0$~eV.
In contrast, the AlO$_2$ layers feature generally the highest formation energies up to $3.3$~eV.
The interaction between the vacancies is largely repulsive,
particularly for compressive strain. % formation energies increase with vacancy concentration.
The first-principles simulations unravel that
the released electrons are accommodated exclusively in the NiO$_2$ layers.
% irrespective of strain and the vacancy location or concentration.
For a vacancy in an AlO$_2$ layer, charge transfer % to the NiO$_2$ layers
spatially separates the donor and the doped NiO$_2$ layers
and reduces the formation energy by up to $70\%$ with respect to bulk LAO.
For a vacancy in a NiO$_2$ layer, the adjacent Ni magnetic moments are enhanced, % in line with Hund's rule,
at variance with bulk LNO.
Site disproportionation proves to be robust for vacancy concentrations up to $0.5\%$,
but is generally quenched for higher values.
By tracking the evolution of the electronic structure with increasing vacancy concentration 
in a NiO$_2$ layer at tensile strain, % from $0\%$ to $8.3\%$,
we observe an unexpected transition from a localized ($0.5\%$) to a delocalized ($2.1\%$) charge accommodation,
a MIT despite a half-filled majority-spin Ni~$e_g$ manifold ($4.2\%$), % a delocalized charge accommodation in 
and finally a magnetic phase transition ($8.3\%$).
The insulating phase at $4.2\%$ % vacancy concentration
features a band gap of up to $\sim 0.5$~eV at compressive strain
and is due to (re-emerging) site disproportionation
driven by quantum confinement, epitaxial strain, and vacancy-induced charge doping.
At tensile strain, the band gap tends to be smaller or the system retains its metallic phase,
which is in sharp contrast to the defect-free system.
These effects vary distinctly from rigid-band shifts.
This study exemplifies the complex impact of oxygen vacancies on the electronic phase in correlated artificial quantum materials~\cite{Radhakrishnan-Geisler-YVOLAO:21, Radhakrishnan-Geisler-YVOLAO:22}.

\section{Methodology}

We performed spin-polarized density functional theory (DFT) calculations~\cite{KoSh65}
within the plane-wave and ultrasoft pseudopotential framework
as implemented in Quantum Espresso~\cite{Vanderbilt:1990, PWSCF}.
We used cutoff energies for the wave functions and density of $35$ and $350$~Ry, respectively,
and the generalized gradient approximation for the exchange-correlation functional  
as parametrized by Perdew, Burke, and Ernzerhof (PBE)~\cite{PeBu96}.
Static correlation effects were considered within DFT$+U$~\cite{Anisimov:93},
employing $U=4$~eV at the Ni sites,
similar to previous work~\cite{May:10, ABR:11, KimHan:15, Geisler-LNOSTO:17, WrobelGeisler:18, GeislerPentcheva-LNOLAO:18, GeislerPentcheva-LNOLAO-Resonances:19}.
We model the (LNO)$_1$/(LAO)$_1$(001) SLs
with and without explicit oxygen vacancies (\VOs)
by using supercells of increasing size,
simultaneously taking octahedral rotations fully into account:
a $20$-atom $\sqrt{2}a \times \sqrt{2}a \times 2c$ cell
rotated by $45^\circ$ around the pseudocubic $c$~axis,
a $40$-atom $2a \times 2a \times 2c$ cell,
an $80$-atom $2a \times 2a \times 4c$ cell,
and a $320$-atom $4a \times 4a \times 4c$ cell.
These cell sizes correspond to \VO concentrations of $8.3\%$, $4.2\%$, $2.1\%$, and $0.5\%$, respectively. % of the oxygen sites
The Brillouin zone was sampled by using
$12 \times 12 \times 8$,
$8 \times 8 \times 8$,
$8 \times 8 \times 4$,
and $4 \times 4 \times 4$
$\Vek{k}$-point grids~\cite{MoPa76}, respectively,
and $5$~mRy smearing~\cite{MePa89}.
We considered the effect of epitaxial strain
by setting the in-plane lattice parameter to
$a_\text{LAO} = 3.79~\AA$ (compressive strain) or $a_\text{STO} = 3.905~\AA$ (tensile strain).
The out-of-plane lattice parameter was set to literature values, $c = 3.93~\AA$ for $a_\text{LAO}$ and $c = 3.83~\AA$ for $a_\text{STO}$~\cite{ABR:11,Freeland:11}.
For strained LNO, we employed
$c = 3.895~\AA$ for $a_\text{LAO}$ and $c = 3.807~\AA$ for $a_\text{STO}$
to ensure consistency with the literature~\cite{May:10}.
In all cases, the atomic positions were carefully optimized until the residual forces on the ions were below $1$~mRy/a.u.
The formation energies of \VO under oxygen-rich conditions were subsequently calculated from the DFT total energies \footnote{The well-known overbinding of gas-phase O$_2$ molecules in DFT~\cite{LNO-OxVac-Beigi:15} necessitates a correction of $E({\text{O}_2})$, which we performed such as to reproduce the experimental O$_2$ binding energy of $5.16$~eV.} via
\begin{equation*}
  E_{\text{V}_\text{O}}^f = E({\text{system with V}_\text{O}}) - E(\text{ideal system}) + \tfrac{1}{2} E({\text{O}_2}) \ \text{.}
\end{equation*}
%

% The DOS is normalized per formula unit.
% The adjacent broken octahedra are highlighted for clarity. 
\begin{figure}
	\centering
	\includegraphics[width=8.6cm]{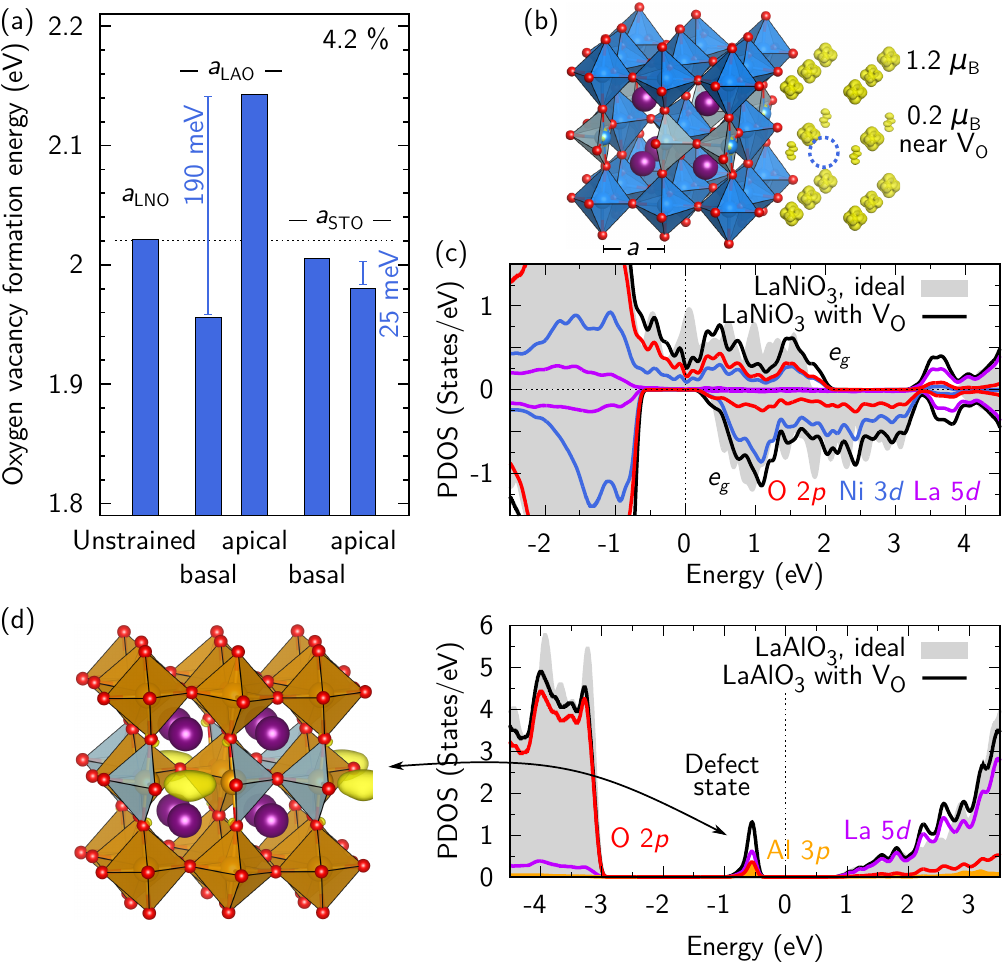}
	\caption{(a)~Emergent anisotropy in the oxygen vacancy formation energies $E_{\text{V}_\text{O}}^f$ in bulk LNO for compressive ($a_\text{LAO}$) and tensile ($a_\text{STO}$) epitaxial strain, relative to the unstrained reference. (b)~The spin density illustrates the accommodation of the released electrons by a collapse of the adjacent Ni magnetic moments, which occurs irrespective of strain. (c)~Spin-resolved PDOS of defect-free LNO versus $4.2\%$ \VO in LNO in the unstrained case.
	(d)~In bulk LAO, $4.2\%$ \VO give rise to a defect state in the band gap that corresponds to an electron accumulation at the \VO site, as illustrated by the charge density integrated between $-1$~eV and $E_\text{F}$. A comparable state can be identified in bulk LNO at $\sim 3.6$~eV [see panel~(c)].}
	\label{fig:FormEn-OxVac-LNO}
\end{figure}

\section{\boldmath Anisotropic oxygen vacancy formation in L\lc{a}N\lc{i}O$_3$ induced by strain}

We first explore the effect of epitaxial strain on the \VO formation in bulk LNO,
considering the exemplary case of $4.2\%$ \VOs.
This serves on one hand as a reference for the (LNO)$_1$/(LAO)$_1$(001) SLs discussed below,
but is additionally motivated by the topical infinite-layer phase of superconducting rare-earth nickelates~\cite{Li-Supercond-Inf-NNO-STO:19, Osada-PrNiO2-SC:20, Li-Supercond-Dome-Inf-NNO-STO:20, Gu-NNO2:20, Lu-MagExNdNiO2:21, Osada-LaNiO2-SC:21},
which is obtained by exclusive reduction of the \textit{apical} oxygen sites towards $33\%$ \VO concentration.
In rhombohedral bulk LNO, the formation energy $E_{\text{V}_\text{O}}^f$ is isotropic.
Figure~\ref{fig:FormEn-OxVac-LNO}(a) shows for this representative nickelate that
strain induces a significant anisotropy in $E_{\text{V}_\text{O}}^f$:
Compressive strain ($a_\text{LAO}$) stabilizes the formation of basal \VOs,
simultaneously rendering a high formation energy for apical \VO with a massive difference of $190$~meV$/$\VOs.
In stark contrast, tensile strain ($a_\text{STO}$) facilitates \VO formation in general,
and particularly inverts the anisotropy in favor of the apical oxygen ions.
While the energy difference is smaller in this case ($25$~meV$/$\VOs),
this mechanism is expected to induce a more controlled reduction to the infinite-layer phase
in epitaxial films as compared to the unstrained bulk,
which offers a possible explanation for the absence of superconductivity in the latter~\cite{Li-NoSCinBulkDopedNNO:19, Wang-NoSCinBulkDopedNNO:20}.
Additional effects are likely related to the polar interface to the STO(001) substrate and the surface:
In film geometry, the reduction energy of the apical oxygen ions
varies from $\sim 0.5$~eV (surface) to $3.5$~eV (interface)~\cite{GeislerPentcheva-InfNNO:20}.
%  (electrostatic doping, electronic reconstruction)

Interestingly, the two electrons released by the \VO are accommodated in the metallic bulk LNO
without changing the total magnetic moment of the supercell ($1~\muB/$Ni).
Instead, we observe a collapse of the local magnetic moment at the two adjacent Ni sites from $\sim 1.0$ to $\sim 0.2~\muB$ (formally Ni$^{2+\delta}$)
and a concomitant slight increase at the six remaining Ni sites from $\sim 1.0$ to $\sim 1.2~\muB$ (Ni$^{3-\delta/3}$) [Fig.~\ref{fig:FormEn-OxVac-LNO}(b)].
Here, $\delta > 0$ accounts for the partial delocalization of the charges.
This mechanism occurs consistently for all degrees of strain considered here.
We see that such local changes are rather indicated by the magnetic moments,
since the local charge varies merely by $\sim 0.15~e^-$ between the different oxidation states,
consistent with previous findings in (LNO)$_1$/(LAO)$_1$(001) SLs~\cite{ABR:11} and other oxides~\cite{OxideValence-Quan:12}.

The projected density of states (PDOS) of reduced LNO [Fig.~\ref{fig:FormEn-OxVac-LNO}(c)] exhibits two notable differences with respect to the defect-free case.
First, it is substantially lowered around $E_\text{F}$,
which indicates that additional effects such as strain or quantum confinement (see below) may drive a MIT.
Second, a La-$5d$-dominated state emerges at $\sim 3.6$~eV,
which is a recurrent motif observed for all reduced systems throughout this work.

\section{\boldmath Oxygen vacancy formation in strained (L\lc{a}N\lc{i}O$_3$)$_1$/(L\lc{a}A\lc{l}O$_3$)$_1$(001) superlattices}

\begin{figure}
	\centering
	\includegraphics[width=8.6cm]{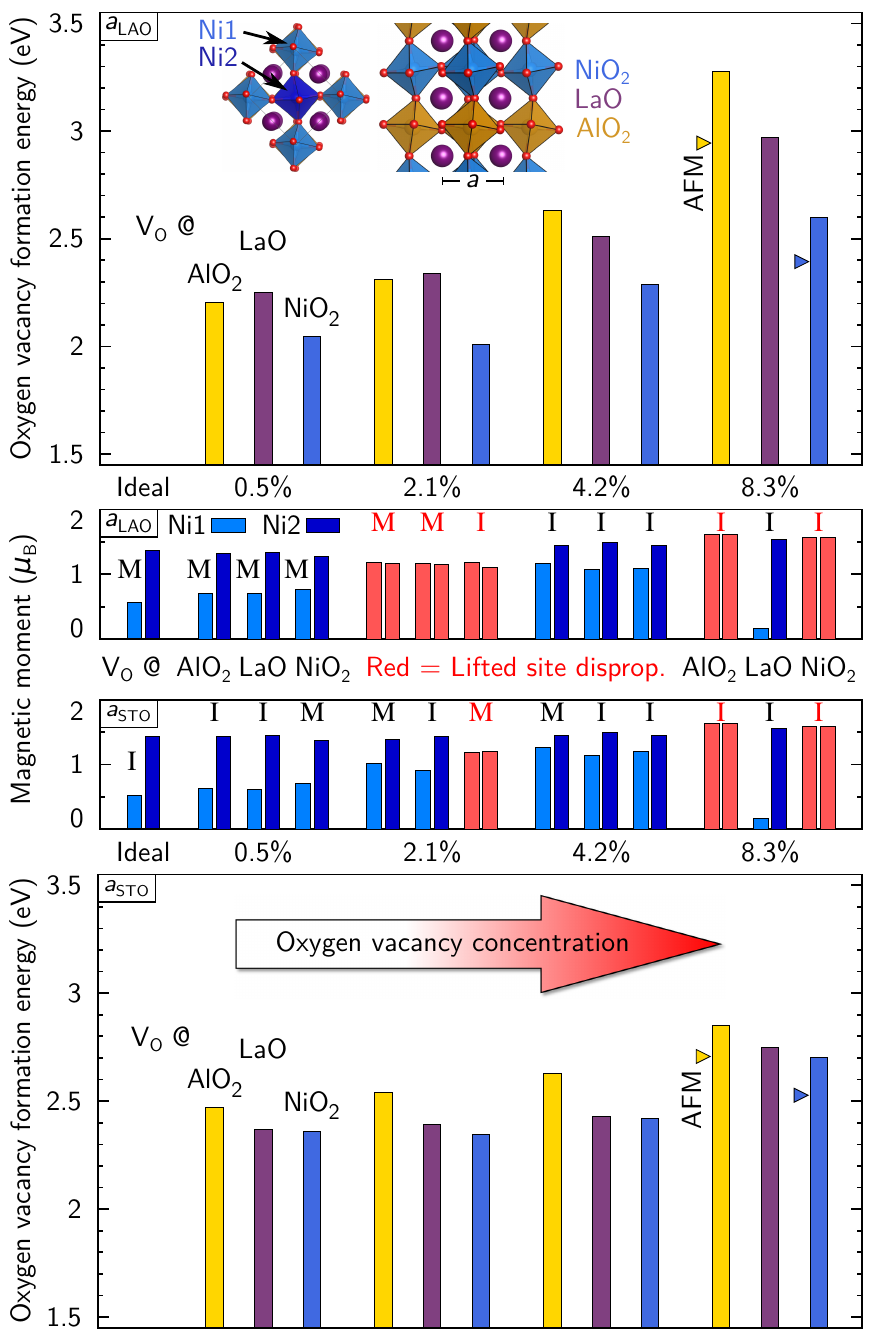}
	\caption{Oxygen vacancy formation energies $E_{\text{V}_\text{O}}^f$ in different layers of (LNO)$_1$/(LAO)$_1$(001) SLs at compressive ($a_\text{LAO}$, top) and tensile ($a_\text{STO}$, bottom) epitaxial strain as a function of the \VO concentration.
	For each case, local Ni1 and Ni2 magnetic moments, averaged over the respective checkerboard sublattices, indicate the persistence (black) or quenching (red) of site disproportionation that is characteristic of the ideal (i.e., defect-free) SL.
	Additionally, it is shown whether the system is metallic~(M) or insulating~(I).
	An overall increasing trend of the Ni magnetic moments with the \VO concentration can be observed.
	For $8.3\%$ \VOs, the system is either rendered AFM (\VO in AlO$_2$ or NiO$_2$ layer; energy marked by triangles) or exhibits one collapsing Ni magnetic moment (\VO in LaO layer).}
	\label{fig:FormEn-OxVac}
\end{figure}

\begin{figure*}
	\centering
	\includegraphics[width=\textwidth]{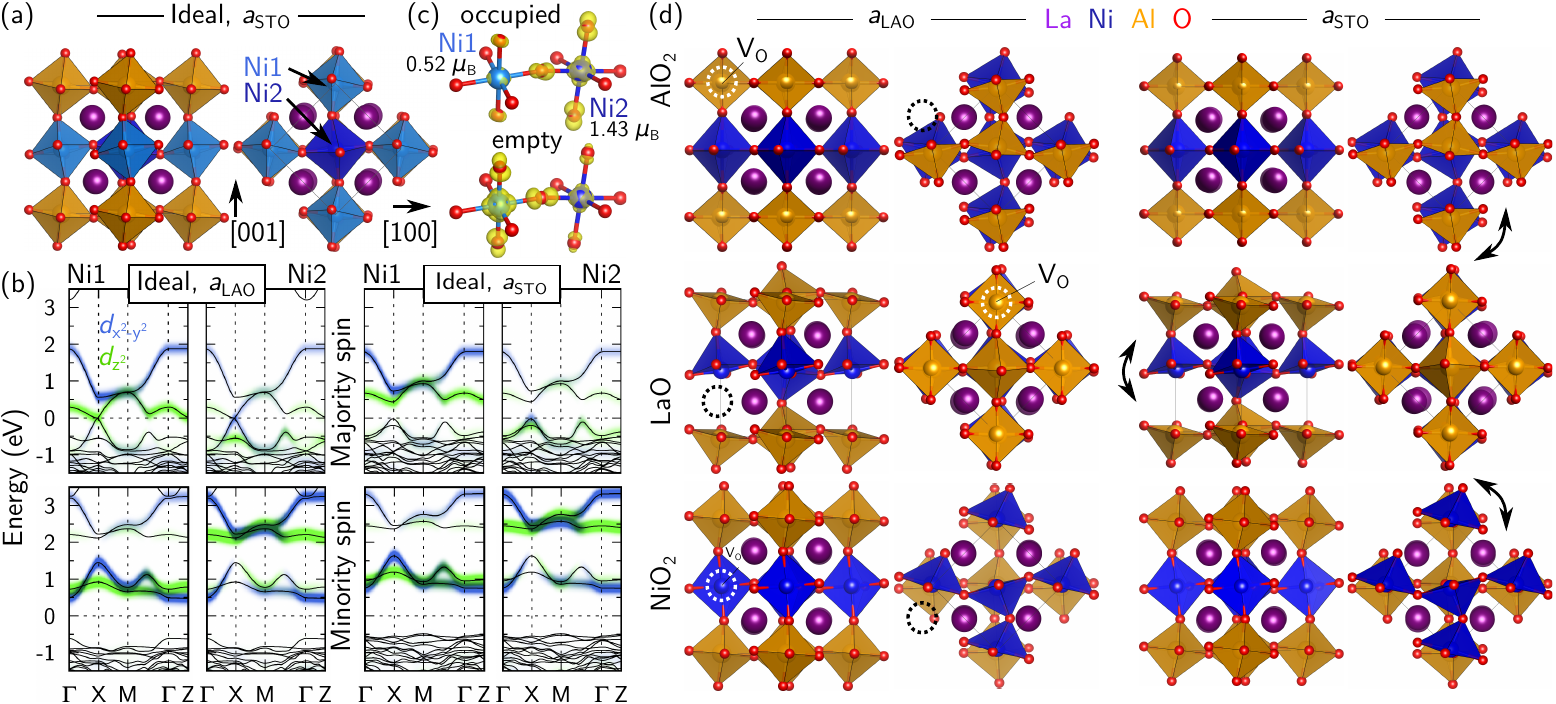}
	\caption{(a)~Side and top view of the optimized ideal (i.e., defect-free) (LNO)$_1$/(LAO)$_1$(001) SL at $a_\text{STO}$. (b)~Spin- and site-resolved band structures show the confinement-induced MIT driven by epitaxial strain in this system. The Ni orbital contributions are highlighted in green ($3d_{z^2}$) and blue ($3d_{x^2-y^2}$). (c)~Majority spin charge densities, integrated at $a_\text{STO}$ between $-0.6$~eV ($+2$~eV) and $E_\text{F}$, illustrate the site disproportionation. (d)~Optimized SL geometries with $8.3\%$ \VO in one of the three inequivalent layers [from top to bottom: AlO$_2$ (basal), LaO (apical), and NiO$_2$ (basal)] at compressive ($a_\text{LAO}$, left) and tensile ($a_\text{STO}$, right) epitaxial strain. Dashed circles indicate the \VO positions, and arrows emphasize the vacancy-enhanced octahedral rotations.}
	\label{fig:Structures20}
\end{figure*}

In sharp contrast to bulk LNO,
(LNO)$_1$/(LAO)$_1$(001) SLs are intrinsically anisotropic and feature three inequivalent layers:
% an AlO$_2$, a LaO, and a NiO$_2$ layer.
AlO$_2$ and NiO$_2$ with basal oxygen ions,
and LaO with apical oxygen ions.
Figure~\ref{fig:FormEn-OxVac} shows \mbox{layer-,} strain-, and concentration-resolved formation energies $E_{\text{V}_\text{O}}^f$ of oxygen vacancies in (LNO)$_1$/(LAO)$_1$(001) SLs.
As in bulk LNO, $E_{\text{V}_\text{O}}^f$ is always positive, i.e., \VO formation is consistently endothermic.
In general, we find that the AlO$_2$ layer exhibits the highest $E_{\text{V}_\text{O}}^f$
% reaching up to $\sim 3.3$~eV (at $8.3\%$, $a_\text{LAO}$),
($2.2$-$3.3$~eV),
whereas the NiO$_2$ layer bears the lowest $E_{\text{V}_\text{O}}^f$
% with a minimum of $\sim 2.0$~eV (at $2.1\%$, $a_\text{LAO}$).
($2.0$-$2.6$~eV).
The LaO layer, where the \VO are apical, % takes on an interesting role.
features intermediate energies for higher \VO concentrations of $8.3\%$ and $4.2\%$ at $a_\text{LAO}$ ($3.0$ and $2.5$~eV, respectively).
For more dilute \VO concentrations of $2.1\%$ and $0.5\%$, the LaO energies exceed the AlO$_2$ values.
% i.e., they are relatively high.
Surprisingly, at $a_\text{STO}$, the energies required to create a vacancy in the LaO layer
are rather similar to the NiO$_2$ values. % irrespective of the concentration. %, i.e., they are relatively low. % ($2.4$-$2.7$~eV).
This strain-induced reversal is reminiscent of the apical vs.\ basal \VO anisotropy in bulk LNO [Fig.~\ref{fig:FormEn-OxVac-LNO}(a)]
and may have implications during the initial reduction phase.

The high formation energies in the AlO$_2$ layer can be related to bulk LAO,
% here = re-calculated for bulk LAO
where $E_{\text{V}_\text{O}}^f = 6.97$~eV at $4.2\%$ ($6.9$~eV in Ref.~\cite{OxVac-STO-LAO-Mitra:12}).
In this material, a \VO does not induce $n$-type doping, but gives rise to a defect state
with mixed La~$5d$--O~$2p$--Al~$3p$ character in the center of the wide band gap [Fig.~\ref{fig:FormEn-OxVac-LNO}(d)]
that corresponds to a strong accumulation of excess charge at the \VO site.
If the vacancy is doubly ionized to V$_\text{O}^{++}$, $E_{\text{V}_\text{O}}^f$ reduces to $\sim 3.5$~eV~\cite{OxVac-STO-LAO-Mitra:12},
which constitutes a strong driving force for charge transfer.
In the SLs, charge transfer to the NiO$_2$ layer prevents the occupation of this defect state, as we will see in the following.
The corresponding state in bulk LNO can be identified at $\sim 3.6$~eV,
as mentioned above [Fig.~\ref{fig:FormEn-OxVac-LNO}(c)].
Since it is empty, $E_{\text{V}_\text{O}}^f$ is much lower in bulk LNO
% here = re-calculated for bulk LNO
($2.02$~eV at $4.2\%$ here; $\sim 2.8$~eV in Ref.~\cite{LNO-OxVac-Beigi:15}).
The $E_{\text{V}_\text{O}}^f$ values for the ultrathin SLs are comparable to
(LNO)$_3$/(LAO)$_1$(001) SLs at $a_\text{STO}$ ($\sim 2.3$~eV~\cite{GeislerPentcheva-LNOLAO-Resonances:19})
as well as to recently reported formation energies of complete apical \VO layers,
i.e., reduction energies from the perovskite to the infinite-layer phase,
which amount to
$2.8$ (LaNiO$_3$),
$2.7$ (PrNiO$_3$),
$2.7$ (NdNiO$_3$), and
$6.7$~eV (LaAlO$_3$)~\cite{SahinovicGeisler:21}.

We observe a strong increase of $E_{\text{V}_\text{O}}^f$ with the \VO concentration (Fig.~\ref{fig:FormEn-OxVac}),
indicative of a repulsive \VOs-\VO interaction.
% namely $8.3\%$, $4.2\%$, $2.1\%$, and $0.5\%$ of the oxygen sites,
This increase is far more pronounced for compressive than for tensile strain;
e.g., in the AlO$_2$ layer,
$E_{\text{V}_\text{O}}^f(8.3\%)-E_{\text{V}_\text{O}}^f(0.5\%) \approx 1.1$~eV ($a_\text{LAO}$) contrasts with $\approx 0.4$~eV ($a_\text{STO}$).
In the highly dilute limit of infinite \VO separation,
$E_{\text{V}_\text{O}}^f$ is higher for tensile than for compressive strain, irrespective of the layer.
For vacancies in the NiO$_2$ layer, the repulsive trend is followed by a slight increase of $E_{\text{V}_\text{O}}^f$ while shifting from $2.1\%$ to $0.5\%$ \VO concentration,
indicative of an onset of attractive behavior.

\section{\boldmath Evolution of the electronic and magnetic phase in (L\lc{a}N\lc{i}O$_3$)$_1$/(L\lc{a}A\lc{l}O$_3$)$_1$(001) superlattices}

\subsection{The defect-free reference system}

Ideal (i.e., defect-free) (LNO)$_1$/(LAO)$_1$(001) SLs [Fig.~\ref{fig:Structures20}(a)] exhibit a confinement-induced MIT driven by strain~\cite{ABR:11, Freeland:11}:
being still metallic at $a_\text{LAO}$, a band gap of $\sim 0.3$~eV opens for tensile strain [Fig.~\ref{fig:Structures20}(b)]
due to a disproportionation at the Ni sites [Fig.~\ref{fig:Structures20}(c)],
expressed in a checkerboard-like variation of the Ni1$/$Ni2 magnetic moments,
% Simon
  $0.56/1.36~\muB$ ($a_\text{LAO}$) and $0.52/1.43~\muB$ ($a_\text{STO}$),
and accompanied by substantial breathing-mode structural distortions of the Ni octahedra,
featuring Ni1$/$Ni2 volumes of $9.4/10.6$ ($a_\text{LAO}$) and \smash{$9.7/11.2~\AA^3$} ($a_\text{STO}$).
We find ferromagnetic (FM) order to be 135 ($a_\text{LAO}$) and 133~meV$/$Ni ($a_\text{STO}$) more stable than $G$-type antiferromagnetic (AFM) order. %, in line with earlier work~\cite{GeislerPentcheva-LNOLAO:18}.

% Ni2 larger octahedra

The states around $E_\text{F}$ stem from the majority-spin Ni~$e_g$ manifold [Figs.~\ref{fig:Structures20}(b) and~(c)].
% are localized in the NiO$_2$ layers
The valence band maximum exhibits predominantly O~$2p$ and Ni2~$3d$ character,
which explains the higher Ni2 magnetic moment.
In contrast, the conduction band minimum is formed by O~$2p$ states that hybridize with Ni1~$3d_{z^2}$;
at higher energies, Ni1~$3d_{x^2-y^2}$ states can be observed~\cite{ABR:11}.

\subsection{Structural modifications induced by oxygen vacancies}

Figure~\ref{fig:Structures20}(d) shows how \VO affect the crystal structure
of (LNO)$_1$/(LAO)$_1$(001) SLs at $a_\text{LAO}$ and $a_\text{STO}$,
exemplarily displayed in the high-concentration limit ($8.3\%$) to enhance the clarity.
Depending on their location, the impact of \VO varies considerably.
Comparison with the defect-free structure reveals strong modifications
beyond strain effects.
Three major implications can be identified:
(i)~Adjacent octahedra are transformed into pyramids (not tetrahedra).
For Ni, this lowers the energy of the $3d_{z^2}$ orbital (which aligns perpendicular to the pyramidal plane)
relative to $3d_{x^2-y^2}$ (which lies in the pyramidal plane).
(ii)~We find substantial changes 
of the octahedral rotations, which affect the $B$-O bond lengths and the respective band widths.
For instance, the basal Ni-O-Ni bond angles at $a_\text{STO}$ are modified from $160^\circ$ in the defect-free case to $137$-$148^\circ$ for $8.3\%$ \VO in the NiO$_2$ layer (enhanced rotations),
whereas the apical Al-O-Ni bond angles change from $159^\circ$ to $160$-$166^\circ$ (slightly reduced rotations)
[bottom-right structure in Fig.~\ref{fig:Structures20}(d)].
Further inspection of Fig.~\ref{fig:Structures20}(d) reveals with respect to the ideal case:
a reduction to $a^0$ and enhanced $c^-$ rotations for \VO in AlO$_2$;
$a^-$ rotations similar to the defect-free structure and a reduction to almost $c^0$ for \VO in LaO; and
$a^-$ rotations similar to the defect-free structure and enhanced $c^-$ rotations for \VO in NiO$_2$.
(iii)~We observe elongations of the (intact) Ni octahedra, particularly Ni1, along the $c$ direction.
For instance, $8.3\%$ \VO in the AlO$_2$ layer increase the apical O-Ni1-O distance from $3.93$ to $4.37~\AA$ ($a_\text{LAO}$) and from $3.90$ to $4.28~\AA$ ($a_\text{STO}$).
This correlates with an increased occupation of the Ni1~$e_g$ states
% (initially Ni1~$3d_{z^2}$ for lower \VO concentrations; subsequently Ni1~$3d_{x^2-y^2}$)
and the respective apical O~$2p_z$ states,
which constitute the conduction band in the defect-free case [Figs.~\ref{fig:Structures20}(b), \ref{fig:Structures20}(c), and~\ref{fig:ElStr-Layers}].

\subsection{Layer-resolved accommodation of \boldmath $4.2\%$ oxygen vacancies}

\begin{figure}
	\centering
	\includegraphics[]{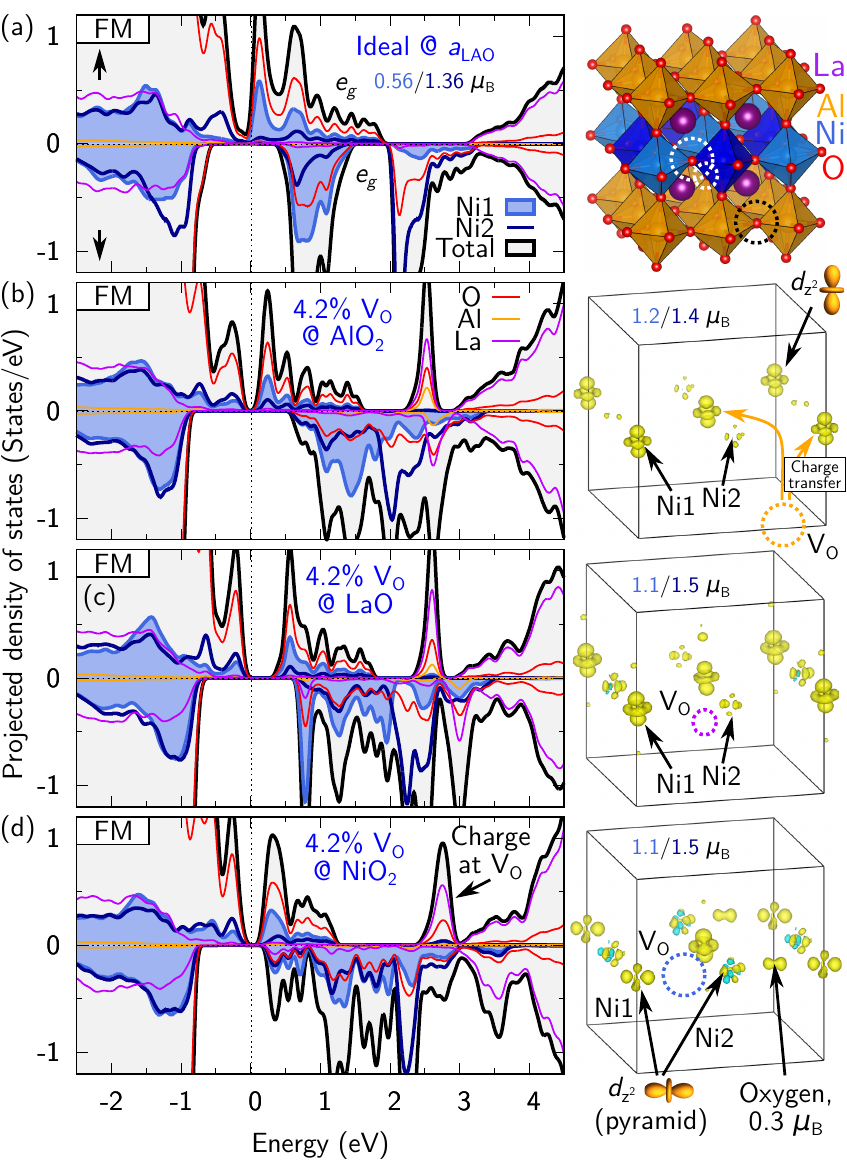}
	\caption{Accommodation of $4.2\%$ \VO located in one of the three inequivalent layers of (LNO)$_1$/(LAO)$_1$(001) SLs at compressive strain ($a_\text{LAO}$).
	The spin- and site-resolved densities of states (left) have been averaged over all corresponding sites in the large supercells and normalized to one SL formula unit.
	The panels on the right show spin density differences with respect to the ideal system (unit cell illustrated at the top), i.e., the vacancy-induced local changes in orbital occupation (yellow: increase; blue: decrease). Irrespective of the \VO location (indicated by dashed circles of corresponding color), the released electrons are accommodated in the NiO$_2$ layer, predominantly by occupation of Ni1~$3d_{z^2}$ orbitals. This is accompanied by an overall enhancement of the Ni1$/$Ni2 magnetic moments.}
	\label{fig:ElStr-Layers}
\end{figure}

\begin{figure}
	\centering
	\includegraphics[]{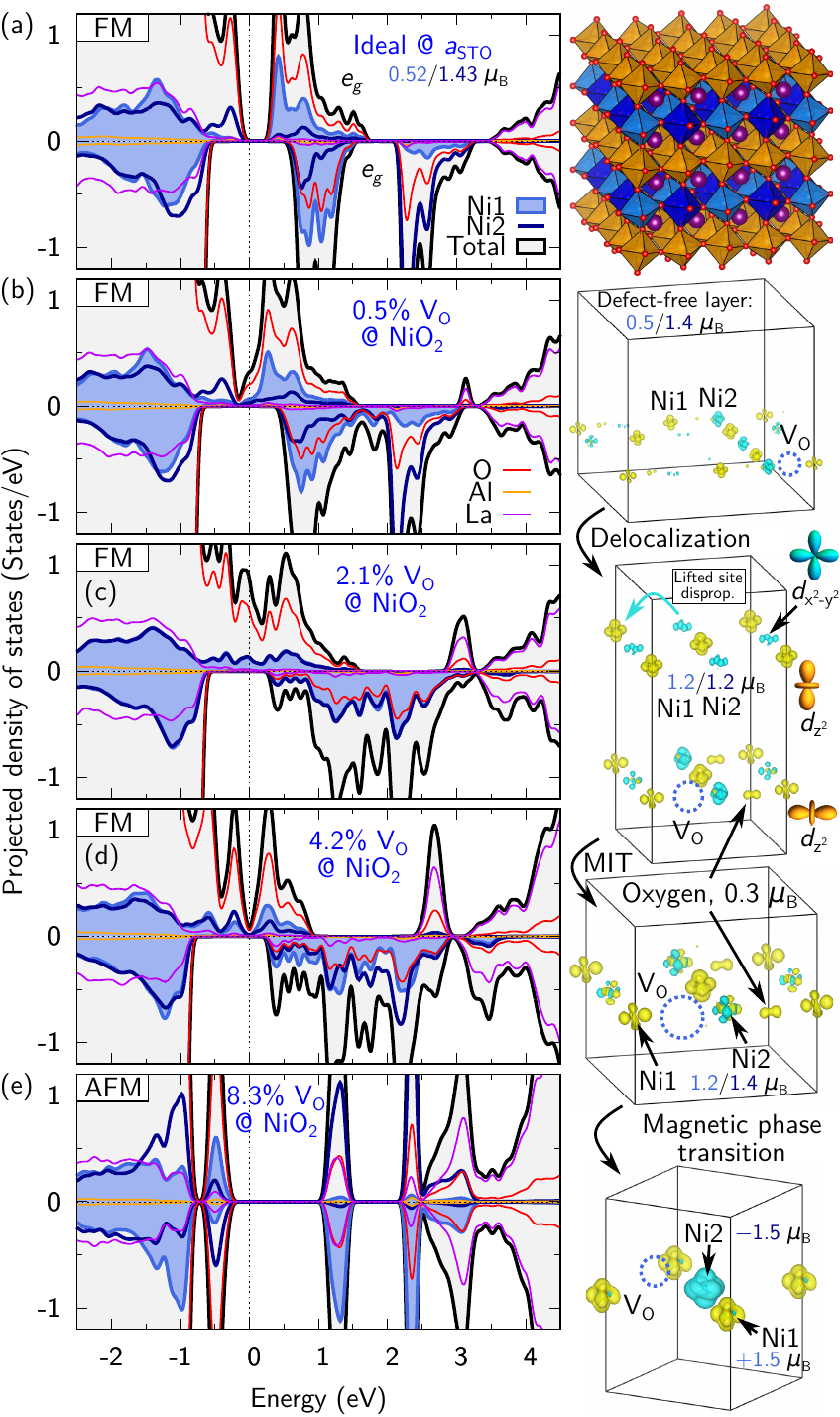}
	\caption{Concentration dependence of the electronic reconstruction in (LNO)$_1$/(LAO)$_1$(001) SLs at tensile strain ($a_\text{STO}$) for oxygen vacancies in the NiO$_2$ layer. With increasing \VO concentration from $0\%$ to $8.3\%$ (i.e., decreasing supercell size, as visible from the spin density difference panels on the right), a localization-delocalization transition, a MIT, and finally a magnetic phase transition are consecutively driven. The increasing Ni1$/$Ni2 magnetic moments reflect the accommodation of the ever growing charge doping.}
	\label{fig:ElStr-NiO2}
\end{figure}

These structural modifications,
which are in part localized at the \VO site,
give rise to a rich variety of accommodation mechanisms
of the two electrons released by each \VOs.
We illustrate these mechanisms in the following, using $4.2\%$ \VO in a SL at $a_\text{LAO}$ as example.
Since the local magnetic moments turn out to be a highly sensitive indicator for changes in the valence state, % the site disproportionation,
we use spin density \textit{differences} in Figs.~\ref{fig:ElStr-Layers} and~\ref{fig:ElStr-NiO2},
$$(n_{\text{SL with V}_\text{O}}^\uparrow - n_{\text{SL with V}_\text{O}}^\downarrow)-(n_\text{Ideal}^\uparrow - n_\text{Ideal}^\downarrow)$$
to spatially trace changes in the electronic reconstruction with respect to the defect-free case.
Figure~\ref{fig:ElStr-Layers} shows that the electrons are accommodated exclusively in the NiO$_2$ layers,
irrespective of the \VO location.
If the \VO is located in an AlO$_2$ layer, the octahedral structure of the NiO$_2$ layers is preserved [Fig.~\ref{fig:Structures20}(d)]. 
The broken Al octahedra give rise to sharp defect states at $\sim 2.5$~eV [Fig.~\ref{fig:ElStr-Layers}(b)] reminiscent of \VO in bulk LAO [Fig.~\ref{fig:FormEn-OxVac-LNO}(d)],
whose charge is transferred to Ni1~$3d_{z^2}$ orbitals pointing in apical direction, as visible in the spin density difference plots [Fig.~\ref{fig:ElStr-Layers}(b)].
Thus, the donor and the doped layer are spatially separated.
If the \VO is located in a LaO layer [Fig.~\ref{fig:ElStr-Layers}(c)], the adjacent Al and Ni octahedra are transformed into pyramids, lowering the energy of the respective Ni~$3d_{z^2}$ orbital (still pointing in apical direction),
similar to the infinite-layer rare-earth nickelates~\cite{Nomura-Inf-NNO:19, Botana-Inf-Nickelates:19, GeislerPentcheva-InfNNO:20, GeislerPentcheva-NNOCCOSTO:21}.
If the \VO is located in a NiO$_2$ layer [Fig.~\ref{fig:ElStr-Layers}(d)], the two adjacent Ni octahedra are transformed into pyramids.
At these sites, the $3d_{z^2}$ orbitals are oriented in the \textit{basal} plane, % and subsequently occupied.
as it can be clearly observed at the Ni1 site.
This orbital reconstruction induces a yellow-blue signature (increase in the basal plane, decrease in the apical direction)
in the spin density difference also at the Ni2 site.
Simultaneously, the oxygen ion that connects the Ni pyramid apexes gets strongly spin-polarized ($0.3~\muB$),
indicative of the covalent nature of the bond [see also Fig.~\ref{fig:Structures20}(c)].
In sharp contrast to bulk LNO [Fig.~\ref{fig:FormEn-OxVac-LNO}(b)], % at variance
here the adjacent Ni magnetic moments do not collapse;
instead, the Ni1 moments are almost doubled ($\sim 1.1~\muB$),
while the Ni2 moments ($\sim 1.5~\muB$) remain close to the ideal case
despite the structural relaxations and orbital reconstructions.

Surprisingly, a band gap of up to $\sim 0.5$~eV emerges at $E_\text{F}$ irrespective of the \VO position at $a_\text{LAO}$ (Fig.~\ref{fig:ElStr-Layers}).
While already bulk LNO showed tendencies towards a vacancy-driven MIT, as discussed above [Fig.~\ref{fig:FormEn-OxVac-LNO}(c)],
in the SL this substantial effect cannot be anticipated from a rigid-band picture,
which would suggest metallic behavior instead % for $4.2\%$ \VO
as the unoccupied Ni $e_g$ bands of the ideal SL are intertwined, irrespective of strain [Fig.~\ref{fig:Structures20}(b)].
At $a_\text{STO}$, this vacancy-induced MIT is less pronounced or absent:
a \VO in a LaO or NiO$_2$ layer induces a band gap of $\sim 0.1$~eV,
whereas the system stays metallic for a \VO in an AlO$_2$ layer [see Fig.~\ref{fig:ElStr-NiO2}(d) and Supplemental Material~\cite{SuppMat}].
This strain dependence is in sharp contrast to the physics of the ideal SL.
Overall, each system at $4.2\%$ \VO retains the site-disproportionated phase.
Thus, we conclude that the MIT arises due to site disproportionation
as a concerted effect of quantum confinement, epitaxial strain, and charge doping by the oxygen vacancies.

\subsection{Evolution of the electronic structure between \boldmath $0\%$ and $8.3\%$ oxygen vacancy concentration}

Motivated by this observation,
we now explore in more detail how the electronic structure of (LNO)$_1$/(LAO)$_1$(001) SLs evolves with increasing \VO concentration.
We focus on tensile strain ($a_\text{STO}$) and a vacancy in a NiO$_2$ layer,
which are most prone to \VO formation, as we discussed above (Fig.~\ref{fig:FormEn-OxVac}).
The ideal system exhibits a band gap of $\sim 0.3$~eV,
and the conduction band minimum is formed by majority-spin Ni1~$3d_{z^2}$--O~$2p$ hybrid states [Figs.~\ref{fig:Structures20}(b) and~\ref{fig:ElStr-NiO2}(a)].
In a rigid-band picture, introducing $0.5\%$ \VO would lead to a slight occupation of these states,
delocalized across all Ni1 sites.
Instead, we observe that the accommodation is localized in a single NiO$_2$ layer
(the supercell contains two repeats in vertical direction)
and clearly does not affect all Ni sites there,
but just within narrow stripes that extend in $a$ and $b$ direction from the \VO [Fig.~\ref{fig:ElStr-NiO2}(b)].
The blue features at some of the Ni2 sites in the spin density differences indicate
either orbital reconstructions [in the pyramids at the \VO site, similar to Fig.~\ref{fig:ElStr-Layers}(d)]
or local alleviations of the site disproportionation (resulting in $\sim 1~\muB$ at the involved Ni sites)
that arise even at intact octahedra
and are promoted by slight, longer-range structural deformations due to the \VOs.
The second, defect-free NiO$_2$ layer retains an identical site disproportionation and Ni magnetic moments as the ideal SL
($\sim 0.5/1.4~\muB$),
and thereby induces no signatures in the spin density difference plots.
The system is metallic in the \VO layer and, apart from local effects, the overall site disproportionation is preserved, % (Fig.~\ref{fig:MagMom},
as corroborated by the substantial difference in Ni1 versus Ni2 PDOS in Fig.~\ref{fig:ElStr-NiO2}(b).
For $0.5\%$ \VO in a LaO or AlO$_2$ layer, the accommodation is maximally confined, involving only the occupation of two adjacent Ni1~$3d_{z^2}$ orbitals
[similar to Figs.~\ref{fig:ElStr-Layers}(b) and~(c)] and rendering an insulating site-disproportionated phase (see Supplemental Material).

% So ist es. Man koennte ja auch sagen, mach eine Lage 4.2%, die andere 0%; aber das passiert nicht.
Surprisingly, if the \VO concentration is increased to $2.1\%$, both NiO$_2$ layers are affected, including the defect-free one [Fig.~\ref{fig:ElStr-NiO2}(c)].
We find that the two electrons released by the \VO are distributed across the entire SL.
Hence, a localization-delocalization transition of the accommodation is observed
as the \VO concentration is increased from $0.5\%$ to $2.1\%$.
The system is metallic and site disproportionation is completely quenched, 
which can bee seen from the perfectly equal Ni1 and Ni2 PDOS in Fig.~\ref{fig:ElStr-NiO2}(c).
The spin density difference of the defect-free NiO$_2$ layer illustrates with particular clarity
that the alleviated site disproportionation
is expressed by a slight reduction of the Ni2~$3d_{x^2-y^2}$ occupation (blue) and a concomitant increase of the Ni1~$3d_{z^2}$ occupation (yellow),
rendering $\sim 1.2~\muB$ at all Ni sites.
This mechanism is equivalent to the local effects arising in the intact octahedra
at $0.5\%$ \VO concentration discussed above [Fig.~\ref{fig:ElStr-NiO2}(b)].
We again observe a substantial magnetic moment at the oxygen ion ($\sim 0.3~\muB$) at the pyramid apex.

Raising the \VO concentration to $4.2\%$ re-establishes site disproportionation,
and a small band gap of $\sim 0.1$~eV opens at $E_\text{F}$, as mentioned above,
despite the delocalized charges [Fig.~\ref{fig:ElStr-NiO2}(d)].

Finally, for $8.3\%$, the majority-spin Ni~$e_g$ bands are completely filled
and the system becomes insulating.
In this situation, site disproportionation 
is no longer favorable.
(Note that the defect-free SL at $a_\text{STO}$ is insulating exclusively due to site disproportionation.)
Moreover, we found that FM order is destabilized, % in this case,
and the system undergoes yet another transition to an insulating AFM phase
with Ni1 and Ni2 magnetic moments of $\pm 1.5~\muB$
and highly localized occupied and empty Ni~$e_g$ states (PDOS peaks at $-0.5$, $1.3$, and $2.4$~eV) [Fig.~\ref{fig:ElStr-NiO2}(e)].
The antiparallel Ni2 magnetic moment is expressed in the strong blue feature in the spin density difference.
(Conventional spin density plots for $8.3\%$ \VO concentration are provided in the Supplemental Material.)
We conclude that the electronic reconstruction in this system is constantly modified while increasing the \VO concentration
from $0\%$ to $8.3\%$,
consecutively driving a localization-delocalization transition ($2.1\%$), a MIT despite a delocalized charge accommodation in a half-filled majority-spin Ni~$e_g$ manifold ($4.2\%$), and finally also a magnetic phase transition ($8.3\%$).

\begin{figure}
	\centering
	\includegraphics[width=8.6cm]{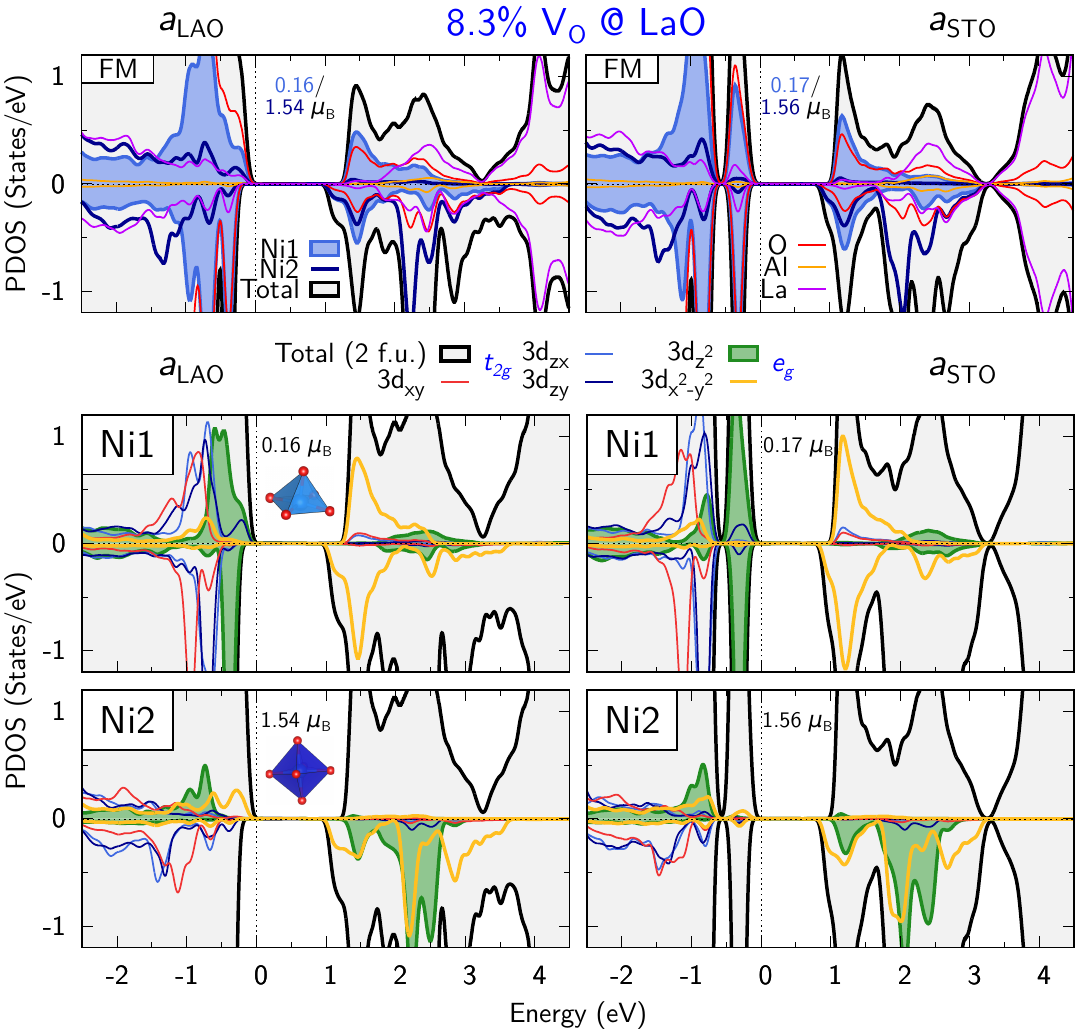}
	\caption{Spin-, site- and orbital-resolved densities of states (top: normalized per f.u.; bottom: original scale) of (LNO)$_1$/(LAO)$_1$(001) SLs for $8.3\%$ \VO in the LaO layer at compressive (left) and tensile (right) epitaxial strain. Analysis of the distinct magnetic moments at the Ni1 and Ni2 sites reveals a different occupation of the $3d_{z^2}$ (green) versus $3d_{x^2-y^2}$ (orange) states resulting from the competition between crystal-field effects and Hund's rule.}
	\label{fig:LaO-PDOS}
\end{figure}

Intriguingly, the localization-delocalization transition at $2.1\%$ is observed irrespective of the \VO location and strain (see Supplemental Material).
The insulating phase at $4.2\%$ [Figs.~\ref{fig:ElStr-Layers} and~\ref{fig:ElStr-NiO2}(d)]
emerges for almost all combinations of \VO location and strain, as discussed above,
with the exception of a \VO in the AlO$_2$ layer. % (where \VO formation is unfavorable anyhow).
Similarly,
the magnetic phase transition at $8.3\%$ \VO concentration
towards AFM order with Ni magnetic moments of $\pm 1.5~\muB$
occurs not only for a \VO in the NiO$_2$ layer, but also for a \VO in the AlO$_2$ layer,
irrespective of epitaxial strain.
The corresponding formation energies are marked by blue triangles in Fig.~\ref{fig:FormEn-OxVac}.
Interestingly, for a \VO in the LaO layer,
we observe a different modification of the magnetic state, as detailed in Fig.~\ref{fig:LaO-PDOS}:
Near the \VOs, the $3d_{z^2}$ orbital is lowered in energy due to the modified crystal field
and doubly occupied (similar to infinite-layer rare-earth nickelates),
leaving the $3d_{x^2-y^2}$ orbital empty and leading to an almost vanishing magnetic moment ($\sim 0.2~\muB$),
whereas in the intact octahedron Hund's rule is followed and the $3d_{z^2}$ and $3d_{x^2-y^2}$ orbitals
are each singly occupied by majority-spin electrons ($\sim 1.55~\muB$).
This mechanism bears analogies to bulk LNO discussed above (Fig.~\ref{fig:FormEn-OxVac-LNO}),
but arises in (LNO)$_1$/(LAO)$_1$(001) SLs exclusively for the highest explored vacancy concentration
(see Supplemental Material for the full data).

We compile these results together with the formation energies in Fig.~\ref{fig:FormEn-OxVac}
to obtain an integrated perspective at the relative stability of the different phases
correlated with their electronic and magnetic properties.
By comparing the local Ni1 and Ni2 magnetic moments,
averaged over the respective sites of the bipartite lattice,
it is shown that various combinations of metallic versus insulating and site-disproportionated versus non-disproportionated phases arise in the SLs depending on the \VO concentration, their location, and the epitaxial strain.
Specifically, the site-disproportionation mechanism is robust up to a \VO concentration of $0.5\%$,
and re-emerges for higher \VO concentrations.
Overall, an increase of the averaged Ni magnetic moments with the \VO concentration can be observed,
which is consistent with the % ever growing fraction of charge accommodated at the Ni sites,
increasing occupation of the Ni~$e_g$ states,
ultimately one additional electron per Ni ion at $8.3\%$.
The lower noninteger Ni magnetic moments of $1.5$-$1.6~\muB$ (formally $2~\muB$) in this case
reflect the covalency in the Ni-O bond, i.e., the strong O~$2p$ involvement
with a finite magnetic moment at the oxygen sites
[Figs.~\ref{fig:Structures20}(c), \ref{fig:ElStr-Layers}, and~\ref{fig:ElStr-NiO2}].

\section{Summary}

We investigated the formation of oxygen vacancies
and their impact on the electronic and magnetic structure
in strained bulk LaNiO$_3$ and (LaNiO$_3$)$_1$/(LaAlO$_3$)$_1$(001) superlattices
by performing comprehensive first-principles simulations including a Coulomb repulsion term.

For bulk LaNiO$_3$, % we demonstrated that 
epitaxial strain induces a substantial anisotropy in the oxygen vacancy formation energy.
Tensile strain as exerted by a SrTiO$_3$(001) substrate enhances the preference for apical vacancy formation by $25$~meV per vacancy,
which would lead to a targeted reduction to the infinite-layer phase
and offers a possible explanation for the exclusive observation of nickelate superconductivity in strained films.
We found that the electrons released by the vacancies are predominantly accommodated locally
despite the overall metallic phase,
resulting in the collapse of the adjacent Ni magnetic moments.
This mechanism is observed irrespective of strain.
Moreover, we identified bulk LaNiO$_3$ at $4.2\%$ vacancy concentration to be at the verge of a vacancy-induced MIT.

Subsequently, we explored the layer and strain dependence of the reduction energies
in (LaNiO$_3$)$_1$/(LaAlO$_3$)$_1$(001) superlattices
and elucidated different accommodation mechanisms of the released electrons,
systematically varying the vacancy concentration between $0\%$ and $8.3\%$.
The density functional theory simulations unraveled that
oxygen vacancies form preferably in the NiO$_2$ layers,
irrespective of strain and the vacancy concentration.
In contrast, the AlO$_2$ layers exhibit generally the highest formation energies.
The reduction is consistently endothermic in an oxygen-rich atmosphere,
and the interaction between the vacancies is largely repulsive,
particularly for compressive strain.
We found that the released electrons are accommodated exclusively in the NiO$_2$ layers.
For a vacancy in an AlO$_2$ layer, this implies a
spatial separation of the donor and the doped NiO$_2$ layers
and reduces the formation energy by up to $70\%$ with respect to bulk LaAlO$_3$.
For a vacancy in a NiO$_2$ layer, the adjacent Ni magnetic moments do not collapse as in bulk LaNiO$_3$;
instead, Hund's rule is followed, leading to enhanced moments
accompanied by a rotation of the respective $3d_{z^2}$ orbitals into the basal plane.
The characteristic site disproportionation is robust for vacancy concentrations up to $0.5\%$,
but generally quenched for higher values.
Detailed analysis of the concentration dependence for the representative case of vacancies in a NiO$_2$ layer at tensile strain
revealed an unexpected transition from a localized ($0.5\%$) to a delocalized ($2.1\%$) charge accommodation,
a metal-to-insulator transition despite a half-filled majority-spin Ni~$e_g$ manifold ($4.2\%$),
and finally also a magnetic phase transition ($8.3\%$).
The insulating phase at $4.2\%$ % vacancy concentration
exhibits a band gap of up to $\sim 0.5$~eV at compressive strain
and is due to (re-emerging) site disproportionation,
which results from the interplay of quantum confinement, epitaxial strain, and vacancy-induced charge doping.
At tensile strain, the band gap is smaller or the system is metallic,
which is in sharp contrast to the defect-free system.
This study exemplifies the rich physics of electronic reconstruction induced by oxygen vacancies in transition-metal oxide heterostructures
that can be captured exclusively by an explicit supercell treatment, % of oxygen vacancies
which points to the limitations of rigid-band approaches
and opens paths for engineering correlated quantum matter.

\section{Acknowledgments}

This work was supported by the German Research Foundation (Deutsche Forschungsgemeinschaft, DFG) within the SFB/TRR~80 (Projektnummer 107745057), Projects No.~G3 and No.~G8.
Computing time was granted by the Center for Computational Sciences and Simulation of the University of Duisburg-Essen
(DFG Grants No.~INST 20876/209-1 FUGG and No.~INST 20876/243-1 FUGG).

\end{document}